\documentclass[twocolumn,10pt]{article}
\usepackage{a4wide,epic}
\usepackage{graphicx}
\usepackage[colorlinks=true]{hyperref}
\usepackage{pstricks}
\usepackage{amsmath}
\usepackage{amssymb}
\usepackage{amsfonts}

\begin{document}

\setlength{\evensidemargin}{-10mm}
\setlength{\oddsidemargin}{-10mm}
\setlength{\textwidth}{180mm}
\setlength{\topmargin}{-10mm}

\title{Non-classical field state stabilization in a cavity by reservoir engineering}

\author{\small A. Sarlette\footnote{alain.sarlette@ulg.ac.be} (Systems \& Modeling, Universit\'e de Li\`ege, B28, 4000 Li\`ege Sart-Tilman, Belgium.)\\ \small J.M. Raimond (Laboratoire Kastler-Brossel, ENS, UMPC-Paris6, CNRS, 24 rue Lhomond, 75005 Paris, France.)\\ 
\small M. Brune (Laboratoire Kastler-Brossel, ENS, UMPC-Paris6, CNRS, 24 rue Lhomond, 75005 Paris, France.)\\
\small P. Rouchon (Centre Automatique et Syst\`{e}mes, Mines ParisTech, 60 boulevard Saint Michel, 75006 Paris, France)}
\date{\today}

\maketitle

\paragraph*{{\large Abstract:}} \textbf{We propose an engineered reservoir inducing the relaxation of a cavity field towards non-classical states. It is made up of two-level atoms crossing the cavity one at a time. Each atom-cavity interaction is first dispersive, then resonant, then dispersive again. The reservoir pointer states are those produced by an effective Kerr Hamiltonian acting on a coherent field. We thereby stabilize squeezed states and quantum superpositions of multiple coherent components in a cavity having a finite damping time. This robust method could be implemented in state-of-the-art experiments and lead to interesting insights into mesoscopic quantum state superpositions and into their protection against decoherence.}\\

Non-classical states of the radiation field are the focus of a considerable interest. Squeezed states (SS), with reduced fluctuations on one field quadrature, are interesting for high precision quantum measurements~\cite{Squeezing}. Mesoscopic field state superpositions (MFSS), involving coherent components with different classical properties, are reminiscent of the famous Schr\"{o}dinger cat~\cite{Schrodinger35}, in a superposition of the ``dead'' and ``alive'' states. Their environment-induced decoherence sheds light on the quantum-classical boundary~\cite{HarocheBook}. We envision in this Letter a reservoir engineering setup in Cavity Quantum Electrodynamics (CQED) for the generation and stabilization of such states.

Many experiments on MFSS have been proposed or realized, particularly with trapped ions~\cite{Winelandcat} (whose harmonic motion is equivalent to a field mode) or CQED~\cite{HarocheBook}, with a single atom coupled to a single field mode. Introducing the atom in a state superposition and finally detecting it leads to the preparation of a MFSS, conditioned by the atomic detection outcome~\cite{HarocheBook,Brune92,Brune96,Deleglise08,Davidovich93,VillasB03,SolanoWalther03,deMatosFilho96}.

Deterministic preparation of MFSS could, in principle, be achieved by propagation of a coherent field in a Kerr medium~\cite{YurkeStoler86}, described by the Hamiltonian:
\begin{equation}\label{eq:Hkerr}
H_{K} = \zeta_K \, \text{\bf{N}} \, +\, \gamma_K \, \text{\bf{N}}^2\ 
\end{equation}
($\text{\bf{N}}$: photon number operator ; $\zeta_K $ is proportional to the linear index ; $\gamma_K$: Kerr frequency; units are chosen such that $\hbar=1$ throughout the paper). An initial coherent state $\vert\alpha \rangle$ evolves with interaction time $t_K$ through nonclassical states $e^{-i\,t_K H_K} \vert \alpha \rangle$ of mean photon number $\vert \alpha \vert^2$~\cite[Section~7.2]{HarocheBook}.
For $t_K\gamma_K \ll \pi$, the field is in a quadrature-squeezed state $\vert s_{\alpha} \rangle$ with a nearly Gaussian Wigner function $W$. For slightly larger interaction times, the field has a `banana'-shaped Wigner function. For $t_K\gamma_K=\pi/k$, we get a MFSS  $\vert k_{\alpha} \rangle$ of $k$ equally spaced coherent components. For $t_K\gamma_K=\pi/2$, a ``Schr\"odinger cat'' state $ \vert c_{\alpha} \rangle = ( \vert \alpha e^{-i\varphi} \rangle + i \, \vert \text{-}\alpha e^{-i\varphi}\rangle)/\sqrt 2$ is reached ($\varphi=\zeta_K t_K$). Note that the collisional interaction Hamiltonian for an atomic sample in a tightly confining potential or in an optical lattice is similar to $H_K$~\cite{Blochcat}.

The unconditional \emph{preparation}, \emph{protection} and long-term \emph{stabilization} of SS and MFSS is an essential goal for the study and practical use of these states. Reservoir engineering~\cite{Poyatos96,Carvalho01} protects target quantum states by coupling the system to an ``engineered'' bath whose pointer states (stable states of the system coupled to the reservoir)~\cite{Zurek81} include the target. The system is effectively decoupled from its standard environment by its much stronger coupling with the engineered bath.

For trapped ions, reservoirs composed of laser fields can stabilize a subspace containing superpositions of coherent vibrations~\cite{Poyatos96,deMatosFilho96}. However, they do not prevent mixing of states belonging to the stabilized subspace, making it impossible to protect a specific MFSS~\cite[pp.~487-488]{HarocheBook}. A complex reservoir could stabilize a superposition of $n$ phonon number states using $n+2$ lasers~\cite{Carvalho01}. In CQED proposals, reservoirs protect squeezed states~\cite{Werlang08} and entanglement of two field modes~\cite{Pielawa07}. In~\cite{Meystre89}, a reservoir made up of a stream of atoms crossing the cavity stabilizes MFSS (cotangent states). However, the scheme is based on the fragile trapping state condition~\cite{TrappingConditions} (a resonant atom entering the cavity in its upper state undergoes a $2p\pi$ quantum Rabi pulse in an $n$-photon field), and state protection is jeopardized by thermal excitations of the cavity at finite temperature.

We propose a robust method to stabilize SS and MFSS in a realistic CQED experiment. It uses an engineered reservoir made up of a stream of 2-level atoms undergoing a tailored composite interaction with the field, dispersive, then resonant, then dispersive again. The reservoir pointer states are then $\approx e^{-i \,t_K H_{K}} \vert \alpha \rangle$, in which $\alpha$ and $\gamma_Kt_K$ can be chosen at will. With proper interaction parameters, we stabilize in particular the states $\vert s_\alpha \rangle$, $\vert k_\alpha \rangle$ and $\vert c_\alpha \rangle$.

\begin{figure}
\includegraphics[width=\columnwidth]{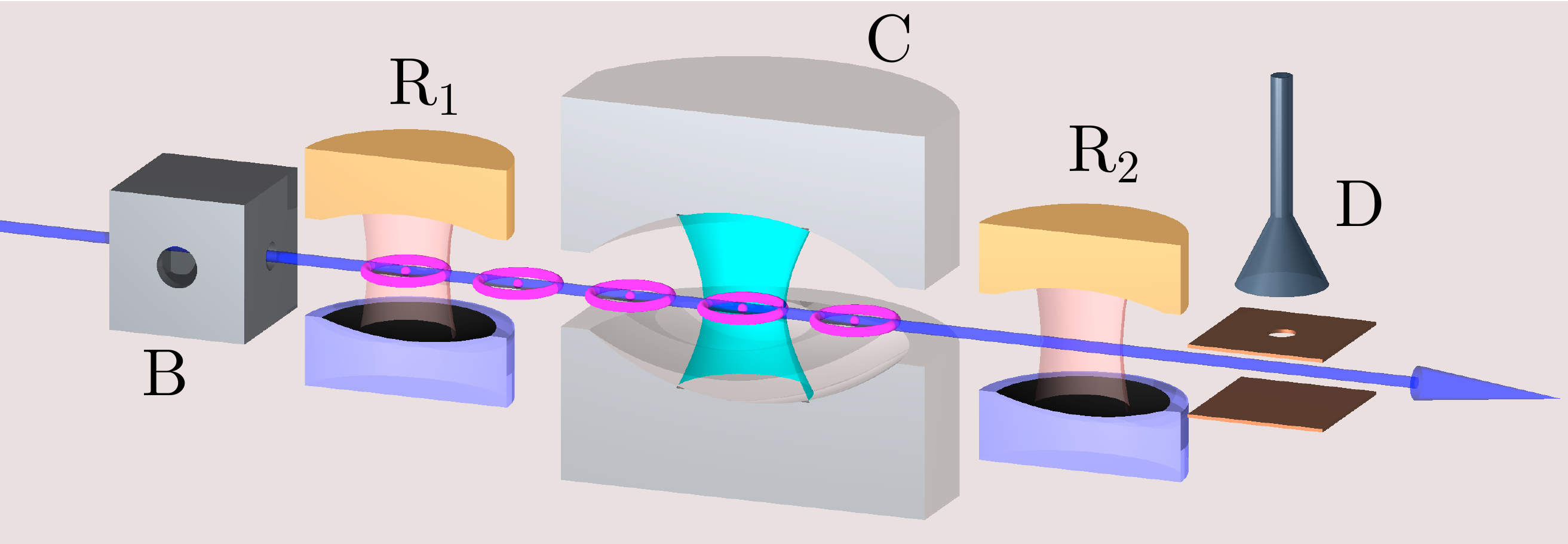}
\caption{Scheme of the ENS CQED experiment.}
\label{fig:ExpLKB}
\end{figure}

The method is quite general and could be applied in a variety of CQED or circuit-QED settings. For the sake of definiteness, we discuss its principle in the context of the ENS CQED setup (Fig.~\ref{fig:ExpLKB}, details in~\cite{HarocheBook,Deleglise08} and below). A microwave field at frequency $\omega_c/2\pi$ is trapped in the cavity $C$. Atoms are sent one after the other through $C$. The transition frequency between the atomic lower and upper circular Rydberg states ($\vert g \rangle$ and $\vert e \rangle$ respectively) is $\omega_0/2\pi$. The atom-cavity detuning $\delta=\omega_0-\omega_c \ll \omega_c$ can be controlled with a good time resolution via the Stark effect. The atoms are excited in state $\vert g \rangle$ in $B$ and prepared by a classical microwave pulse in the Ramsey zone $R_1$ in the initial state $\vert u_\text{at} \rangle = \cos(u/2) \vert g \rangle +  \sin(u/2) \vert e \rangle$ (without loss of generality, we take phase references so that $\langle g\vert u_\text{at} \rangle$ and $\langle e\vert u_\text{at} \rangle$ are both real). On a Bloch sphere with $\vert e \rangle$ at the north pole, it corresponds to a vector at an angle $u$ with the north-south vertical axis. For the engineered reservoir, the final atomic state is irrelevant. Ramsey zone $R_2$ and the state-selective detector $D$ are used to reconstruct the field state generated by the engineered reservoir, using a method described in~\cite{Deleglise08}.

Atom-cavity interaction is ruled by the Jaynes-Cummings Hamiltonian $H_{JC}$. In a proper interaction representation: 
\begin{equation}\label{eq:JCm}
  H_{JC} = \frac{\delta}{2}(\vert e \rangle \langle e \vert-\vert g \rangle\langle g \vert) + i\frac{\Omega(t)}{2} (\, \vert g \rangle \langle e \vert\, \text{\bf{a}}^{\dagger} - \vert e \rangle \langle g \vert\,  \text{\bf{a}} \, )
\end{equation}
where $\Omega(t)$ is the atom-cavity coupling, varying with time during the atomic transit through the Gaussian mode ($\text{\bf{a}}$: photon annihilation operator). The unitary evolution operators corresponding to resonant ($\delta=0$) and strongly dispersive ($|\delta| \gg \Omega$) interactions are, within irrelevant phases~\cite{HarocheBook}:
\begin{eqnarray}
\nonumber 	
\!\! &U_r(\Theta)&
	 =  \vert g \rangle \langle g \vert \, \cos({\scriptstyle \Theta \sqrt{\text{\bf{N}}}}/{\scriptstyle 2})
		  + \vert e \rangle \langle e \vert \, \cos({\scriptstyle \Theta \sqrt{\text{\bf{N}}+\mathbf{1}}}/{\scriptstyle 2}) \\
\label{OnResInt}
\!\! & & - \vert e \rangle \langle g \vert \,\text{\bf{a}} \frac{\sin({\scriptstyle \Theta \sqrt{\text{\bf{N}}}}/{\scriptstyle 2})}{{\scriptstyle \sqrt{\text{\bf{N}}}}} \,
		   +	\vert g \rangle \langle e \vert\, \frac{\sin({\scriptstyle \Theta \sqrt{\text{\bf{N}}}}/{\scriptstyle 2})}{{\scriptstyle \sqrt{\text{\bf{N}}}}} \text{\bf{a}}^{\dagger}\ , \\
\label{OfResInt}
\!\! &U_d(\phi_0)&
	\approx   \vert g \rangle \langle g \vert\;  e^{-i \, \phi_0 \text{\bf{N}}}
	  + \vert e \rangle \langle e \vert\;  e^{+i \, \phi_0 (\text{\bf{N}}+\mathbf{1})} \; ,
\end{eqnarray}
where $\Theta = \int \, \Omega(t) \, dt$ is the quantum Rabi pulsation in vacuum integrated over time during the resonant interaction and $\phi_0 = -1/(4\, \delta)\, \int \Omega^2(t)\,dt$ measures the total field phase shift produced by the atom during the dispersive interaction.

Resonant atoms initially in $|g\rangle$ absorb photons. The reservoir's pointer state is then the vacuum $\vert 0 \rangle$. Resonant atoms initially in $\vert u_\text{at} \rangle$ realize a `micromaser' with coherent injection~\cite{Meystre89,Casagrande02}. Noticeably, when pumped below population inversion ($0<u\ll 1$), with $\Theta\ll 1$, this maser stabilizes a coherent state $\vert \alpha \rangle$. Starting e.g.~from $\vert 0 \rangle$ in an ideal cavity, repeated atomic emissions produce a coherent state with a growing real amplitude $\beta$. The atoms then undergo a coherent resonant Rabi rotation in this field, with a Bloch vector starting initially towards the south pole of the Bloch sphere. Assuming $\Theta\ll u$, the atomic Bloch vector rotates under $U_r(\Theta)$ by an angle $-\Theta\beta$. For $\Theta\beta<2u$, the atomic energy decreases on the average and $\beta$ grows. When $\beta$ reaches $\alpha=2u/\Theta$, the average atomic energy is unchanged after interaction and the field amplitude remains constant as an equilibrium is reached. This intuitive approach is supported, for $u,\Theta\ll 1$, by developing to second order in $u,\Theta$ the master equation for the coarse-grained average of the field density operator map $\rho\rightarrow Tr_{at}\left[U_r(\Theta)\,\left(\rho\otimes|u_{at}\rangle\langle u_{at}|\right)\, U_r(\Theta)^\dagger\right]$~\cite{Tbp1}: the atoms act on the field approximately like a coherent injection plus damping.  For any initial condition, numerical simulations for $u,\Theta\approx 1$ show that the field state rapidly converges towards a pure state close to $|\alpha\rangle$ with $\alpha\approx 2u/\Theta$.

Our key observation is that sandwiching $U_r(\Theta)$ between two opposite dispersive interactions $U_d(\phi_0)$ and $U_d(\phi_0)^\dagger = U_d(-\phi_0)$ is equivalent to sandwiching it between adjoint evolution operators generated by the Kerr Hamiltonian $H_K$. This yields a pointer state resulting from the action of $H_K$ on the pointer state of the resonant reservoir i.e. the coherent state $\vert \alpha \rangle$. 

Using  $\text{\bf{a}} f(\text{\bf{N}})=f(\text{\bf{N}}+\mathbf{1})\, \text{\bf{a}}$, we get for the total evolution operator $U_t=U_d(\phi_0)U_r(\Theta)U_d^\dagger(\phi_0)$:
\begin{eqnarray}
\nonumber 	
U_t & = & \vert g \rangle \langle g \vert \, \cos({\scriptstyle \Theta \sqrt{\text{\bf{N}}}}/{\scriptstyle 2})
		  + \vert e \rangle \langle e \vert \, \cos({\scriptstyle \Theta \sqrt{\text{\bf{N}}+\mathbf{1}}}/{\scriptstyle 2}) \\
\nonumber \!\! & & - \vert e \rangle \langle g \vert \,\text{\bf{a}} \frac{\sin({\scriptstyle \Theta \sqrt{\text{\bf{N}}}}/{\scriptstyle 2})}{{\scriptstyle \sqrt{\text{\bf{N}}}}} \, e^{2i\phi_0\text{\bf{N}}}\\
\label{eq:Ut}		  & & + \vert g \rangle \langle e \vert\, \frac{\sin({\scriptstyle \Theta \sqrt{\text{\bf{N}}}}/{\scriptstyle 2})}{{\scriptstyle \sqrt{\text{\bf{N}}}}} \, e^{-2i\phi_0\text{\bf{N}}} \, \text{\bf{a}}^{\dagger}\ .
\end{eqnarray}
When the atom remains in the same state, the dispersive interactions cancel. When it switches level in $U_r(\Theta)$, the dispersive phase shifts add up. In semi-classical terms, an atomic absorption (emission) decreases (increases) the field amplitude and increases (decreases) its phase. After many atomic interactions, a larger field is expected to have a smaller accumulated phase than a smaller field, in close analogy with the Kerr effect action.

More precisely, using 
$$e^{-i\phi_0\text{\bf{N}}(\text{\bf{N}}+\mathbf{1})}\,\text{\bf{a}}\,e^{i\phi_0\text{\bf{N}}(\text{\bf{N}}+\mathbf{1})}=\text{\bf{a}}\,e^{2i\phi_0\text{\bf{N}}}\ ,$$ and defining $h_0(\text{\bf{N}})=\phi_0\text{\bf{N}}(\text{\bf{N}}+\mathbf{1})$, we have $U_t=\exp[-ih_0(\text{\bf{N}})]\,U_r(\Theta)\,\exp[ih_0(\text{\bf{N}})]$. Since $\exp[-ih_0(\text{\bf{N}})]$ is the evolution operator in $H_K$ for $\gamma_K t_K=\zeta_K t_K=\phi_0$, we get as a pointer state any of the non-classical states produced by $H_K$ by adjusting the interaction parameters. In particular, for $\phi_0 = {\pi}/{2}$, we get, up to an irrelevant phase, the MFSS $\vert c_\alpha \rangle=( \vert \text{-} i\alpha \rangle + i \, \vert i\alpha \rangle)/\sqrt 2$.

Let us give an intuitive insight into the stabilization of this cat state $\vert c_\alpha \rangle$. We assume that, before its interaction with an atom, the field is in the state $\vert \psi_0 \rangle =  ( \vert \text{-} i\alpha_0 \rangle + i \, \vert i\alpha_0 \rangle )/\sqrt 2$ with $\alpha_0<\alpha$~\cite{Note2}. The first dispersive interaction entangles atom and field in a mesoscopic quantum superposition, correlating two $\pi$-phase shifted atomic dipoles  $\vert u_{at} \rangle$ and $\vert (\text{-} u)_{at} \rangle$ with coherent states $\vert \alpha_0 \rangle$ and $\vert \text{-}\alpha_0 \rangle$ respectively. As explained above, during the resonant interaction each dipole state amplifies the correlated coherent component from $\pm \alpha_0$ to $\pm \tilde\alpha$ ($\alpha_0<\tilde\alpha<\alpha$). The second dispersive interaction disentangles atom and field. The final field state is thus independent upon the atomic one and writes $\vert \psi_t \rangle =( \vert \text{-} i\tilde\alpha \rangle + i \, \vert i\tilde\alpha \rangle )/\sqrt 2$, a `larger' MFSS. Similarly, if $\alpha_0>\alpha$, the atomic interaction reduces the cat amplitude. Altogether, the atoms stabilize a sizable MFSS.

Eq.(\ref{OfResInt}) is valid in the large detuning limit. For smaller $\delta$ values, of the order of the maximum atom-field coupling, a more complex expression using the full dressed states should be used. In particular, there is a finite transition probability between atomic levels at the end of the first dispersive interaction. Further analysis~\cite{Tbp1} and numerical simulations of the exact dynamics surprisingly show that our main findings remain quite valid even in this regime. The reservoir's pointer state is still that produced by the Kerr hamiltonian acting onto a coherent state close to $\vert \alpha \rangle$, within a classical phase rotation.

We have performed numerical simulations of the field evolution in realistic experimental conditions, corresponding to the  present ENS CQED setup (Fig.~\ref{fig:ExpLKB}). The cavity and atomic frequencies are close to 51~GHz, corresponding to the transition between circular rubidium Rydberg levels with principal quantum numbers 51 and 50. Taking the origin of time when the atom crosses cavity axis, $\Omega(t) = \Omega_0\; e^{-v^2t^2/w^2}$ in the Gaussian mode of $C$, where $v$ is the adjustable atomic velocity, $\Omega_0/2\pi = 50$~kHz and $w=6$~mm. The simulations take into account the standard cavity relaxation towards thermal equilibrium. The photon lifetime in $C$ is  $T_c = 0.13$~s and the residual thermal field at the mirrors temperature (0.8~K) corresponds to a mean number of blackbody photons per mode $n_t=0.05$. The engineered reservoir is meant to protect non-classical states from this decoherence.

The composite atom-field interaction is achieved with a ladder of Stark shifts during the atom-cavity interaction time. The corresponding evolution operators are computed exactly from $H_{JC}$ [Eq. (\ref{eq:JCm})], using the quantum optics package for MATLAB~\cite{Quopackage} (Hilbert space is truncated to the 60 first Fock states). The atom-field interaction is supposed to start when $vt=-1.5\,w$ (dispersive coupling equal to $\sim 1$\% of its maximum value) and ends when $vt=1.5\,w$, the total interaction time being $t_i=3w/v$. During $t_i$, $\delta$ is first set at $\delta = \Delta > 0$, implementing $U_d^\dagger$, then at $\delta=0$ for a short time span $t_r$ centered on cavity center crossing time, and finally $\delta= -\Delta$ for $U_d$. 

Atomic samples are produced at regular time intervals in $B$. The probability for having one atom in a sample, $p_{at}\simeq 0.3$, is kept low to avoid having two atoms at a time in $C$. The interaction with the next sample starts immediately after the previous one has left $C$. Thus, smaller $t_i$ implies more frequent atom-cavity interaction, that is a stronger engineered reservoir, more efficient against standard cavity relaxation. We trace over the irrelevant final atomic state to compute the field density matrix $\rho$ after each atomic interaction.

\begin{figure*}
\includegraphics[width=\textwidth,trim=0cm 15.8cm 0cm 0cm,clip=true]{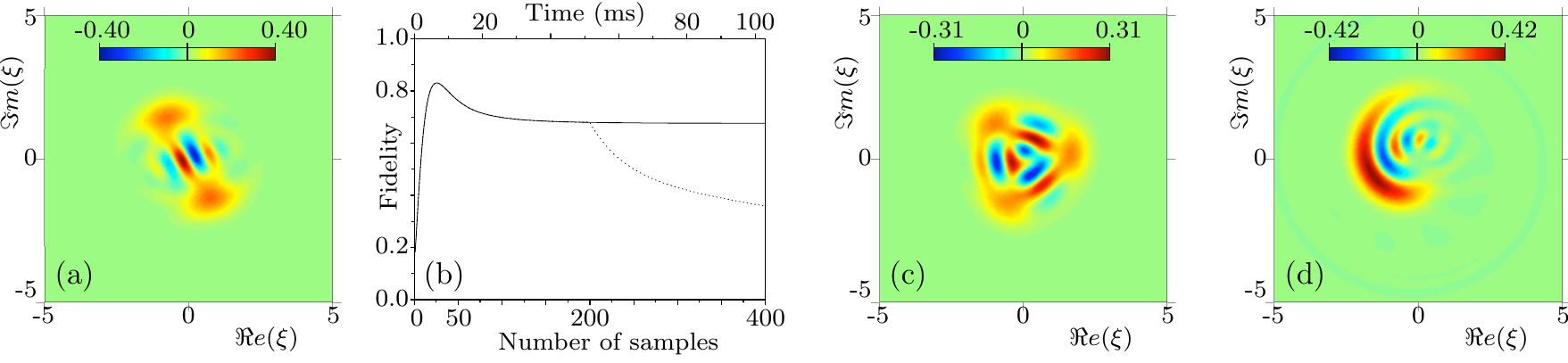}
\caption{Non-classical states generated by the engineered reservoir. (a) Wigner function of the cavity field after 200 steps of reservoir-atom interaction. The state is close to $\vert c_{\alpha} \rangle$. (b) Solid line: fidelity of the generated state against the closest ideal cat as a function of the number of interactions (bottom axis) or of time (upper axis). Dashed line: reservoir is switched off after 200 interactions. (c) and (d) Wigner functions of cavity fields close to $\vert k_{\alpha} \rangle$ with $k=3$ and to a `banana state' respectively. Detailed conditions in the text.}\label{fig:Wig1}
\end{figure*}

Figure \ref{fig:Wig1}(a) presents the experimentally accessible~\cite{Deleglise08} Wigner functions $W(\xi)$ associated to the cavity field state $\rho_{200}$ after its interaction with 200 atomic samples for $v=70$~m/s (requiring a moderate laser cooling of the rubidium atomic beam), $t_i=257\, \mu$s, $\Delta=2.2\,\Omega_0$, $t_r=5\ \mu$s, $\Theta\approx\pi/2$, $u=0.45 \pi$. The (irrelevant) initial cavity state is the vacuum. We get a two-component MFSS with strong negativities  in $W$. The average photon number is $\overline{n}=2.72$. The purity $P=Tr(\rho_{200}^2)$ is 51\%.  We estimate the fidelity $F=Tr[\rho_{200}\rho_{c_\alpha}]$ of this state w.r.t. a MFSS $\rho_{c_\alpha}$ of two coherent states with opposite phases, optimized by adjusting in the reference state the phase and amplitude of the coherent components and their relative quantum phase. We get $F=69$\%. We have checked that cavity relaxation is the main cause of imperfection, $F$ being 98\% in an ideal cavity.

Figure \ref{fig:Wig1}(b) presents as a solid line the fidelity $F$ of the prepared state w.r.t. the final ideal cat as a function of the atomic sample number (i.e. as a function of time). The transient reflects the competition between the fast build-up of the cat, the fidelity raising over a few atomic samples only, and the cat decoherence, proceeding on a slower time scale  $T_d=T_c/(2\overline{n})=27$~ms~\cite{HarocheBook,Deleglise08}. The steady state fidelity is reached after $\simeq 100$ samples. The dashed line presents $F$ when we switch off the reservoir after 200 interactions. It drops much more rapidly than $T_c=0.13$~s, illustrating the high sensitivity of the generated non-classical state to decoherence and its efficient protection by the engineered reservoir.

For a slightly larger detuning, $\Delta=3.7\,\Omega_0$ (all other parameters unchanged), we obtain a three-component MFSS $\vert k_\alpha \rangle$ with $k=3$ [Fig. \ref{fig:Wig1}(c)] with $\overline{n}=2.70$ photons, $P=56\%$, and a fidelity w.r.t. the closest three-component ideal MFSS $F=73\%$.

In the Kerr dynamics, squeezed states are obtained in the early stages of the initial coherent state phase spreading. With $v=300$~m/s, $t_i=60\, \mu$s, $t_r=1.7\ \mu$s i.e.~$\Theta\approx 0.17\pi$, $u=\pi/2$ and $\Delta=70\,\Omega_0$, we generate after 200 samples a Gaussian minimal uncertainty state for the Heisenberg relations between orthogonal field quadratures containing $\overline{n}=21$ photons, with a 1.5~dB squeezing.

For larger phase spreads in the Kerr dynamics, the Wigner function takes a banana shape. These non-minimal uncertainty states present non-classical negativities. As an example, Fig. \ref{fig:Wig1}(d) presents the Wigner function of the field obtained after 200 samples with $v=150$~m/s, $t_i=120\, \mu$s, $t_r=5\ \mu$s i.e. $\Theta\approx\pi/2$, $u=\pi/2$ and $\Delta=7\,\Omega_0$. The field has $\overline{n}=3.52$ and $P=91\%$, due to a reduced influence of relaxation with the fast and more frequent atoms used here. All these settings are within reach of the present ENS setup. A more detailed discussion of these simulations will be published elsewhere~\cite{Tbp1}.

We have checked that this scheme is not sensitive to experimental imperfections (a few \% variation of the interaction parameters does not appreciably modify the steady state), as long as the symmetry between the two dispersive interactions is accurate. It does not require recording the final atomic states, unlike quantum feedback experiments, which can also stabilize non-classical states~\cite{dotsenko-et-al:PRA09}. Feedback can in fact be used in addition to improve the performance of the engineered reservoir, using atomic detection results to detect and react to environment-induced jumps of field state, or to post-select time intervals when a cat is generated with a high fidelity~\cite{Tbp1}.

In conclusion, we have shown that the composite atom-cavity interaction scheme realizes an engineered environment for the cavity field, driving it deterministically towards non-classical field states including the MFSS  Schr\"{o}dinger cat-like states. The scheme is simple and robust enough to be amenable to experiment with a state-of-the-art CQED or circuit QED setup.

The engineered reservoir, driving \emph{any} initial cavity state to the desired mesoscopic field state superposition and stabilizing these quantum resources for \emph{arbitrarily long times}, opens interesting perspectives for fundamental studies of non-classicality and decoherence.

The authors thank I.~Dotsenko, S.~Gleyzes, S.~Haroche and M.~Mirrahimi for enlightening discussions and references. AS is a FNRS postdoctoral visitor at Mines ParisTech and member of the IAP network DYSCO. JMR acknowledges support from the EU and ERC (AQUTE and DECLIC projects) and from the ANR (QUSCO-INCA). PR acknowledges support from ANR (CQUID).


\bibliographystyle{plain}

\end{document}